\documentclass[prb,eqsecnum,twocolumn,floatfix]{revtex4}
\usepackage[dvips]{graphicx}
\usepackage{latexsym}
\usepackage{amsmath}
\usepackage{amssymb}
\begin{document}
\title{Thermal melting of density waves on the square lattice}

\author{Adrian Del Maestro and Subir Sachdev}
\affiliation{Department of Physics, Yale University, P.O. Box
208120, New Haven, CT 06520-8120, USA}

\date{\today}

\begin{abstract}
We present the theory of the effect of thermal fluctuations on
commensurate $p \times p$ density wave ordering on the square
lattice ($p \geq 3$, integer). For the case in which this order is
lost by a second order transition, we argue that the adjacent
state is generically an incommensurate striped state, with
commensurate $p$-periodic long range order along one direction,
and incommensurate quasi-long-range order along the orthogonal
direction. We also present the routes by which the fully
disordered high temperature state can be reached. For $p=4$, and
at special commensurate densities, the $4 \times 4$ commensurate
state can melt directly into the disordered state via a self-dual
critical point with non-universal exponents.
\end{abstract}

\maketitle

\section{Introduction}
\label{sec:intro}

A variety of remarkable recent low temperature scanning tunneling
microscopy (STM) observations have revealed periodic modulations
in the local density of states of the cuprate superconductors
\cite{howald,mcelroy,hanaguri,fang}. Notably, the observations of
Hanaguri {\em et al} \cite{hanaguri} Ca$_{2-x}$Na$_x$CuO$_2$Cl$_2$
clearly show that the modulations have a commensurate period of 4
lattice spacings along both directions of the underlying square
lattice ($4 \times 4$ ordering). Closely related, but not
identical, modulations have been observed in higher temperature
STM \cite{ali}, and in neutron scattering
\cite{jtran,hinkov,buyers,hayden} observations. The differences
between the various experiments relate ({\em i\/}) to the period
of the ordering, which is incommensurate in Ref.~\onlinecite{ali},
and ({\em ii\/}) to whether the ordering extends along one or both
of the $x$ and $y$ axes---neutron scattering experiments are more
easily explained by anisotropic ordering along one of the axes
directions \cite{MVTU,GSU,vs}. An important open question is
whether these differences reflect a fundamental distinction in the
underlying electronic structure of the cuprates, or they can be
explained by the differing experimental parameters of temperature
and carrier concentration.

This paper will begin with the assumption that at very low $T$,
for a small range of carrier densities, and in a sufficiently
clean sample, the system has perfect $4 \times 4$ density wave
order (and, more generally, $p \times p$ order with $p \geq 3$,
integer). We will then present a general phenomenological theory
of the effect of thermal fluctuations on such an ordered state,
describing the phases that appear at higher $T$ and for a wide
range of carrier concentration. Because we focus exclusively on
thermal fluctuations, our results are quite general and
independent of the precise microscopic nature of the density wave
ordering; in particular, the ordering could be a site charge
density wave, or a modulation in exchange or pairing energies due
to valence bond solid order. Our results depend only on the fact
that spin-singlet observables acquire a periodic modulation at low
$T$, and follow completely from the symmetry properties of such
states. The thermal fluctuations of such an ordered state can, and
will, be described by a purely classical theory of the density
wave order parameters.

There is a actually large early literature on the melting of a
variety to two-dimensional solids on different substrates, and on
the commensurate-incommensurate transition
\cite{nh,apy,pt,ostlund,sue,schulz,haldane,huse}. However, most of
this work considered the case of the triangular solid on
substrates with six-fold symmetry, or the
commensurate-incommensurate transition on anisotropic solids. The
particular case of interest here, a square lattice substrate and
isotropic $p \times p$ ordering, appears not to have been
considered previously. We will therefore present an extension of
this earlier theory to the situation appropriate for the cuprate
superconductors. As we will show below, our results have a number
of novel features not found in the cases studied earlier.

\begin{figure}
\centering
\includegraphics[width=3in]{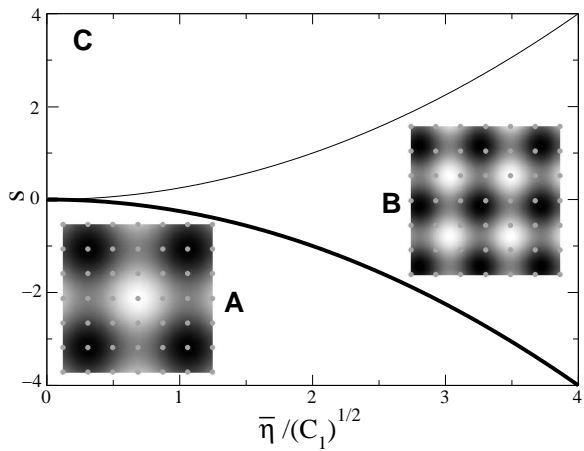}
\caption{Mean field phase diagram of the free energy
$\mathcal{F}_{\Phi}$ in Eq.~(\ref{FP}). Although all parameters in
$\mathcal{F}_{\Phi}$ are functions of both temperature and carrier
concentration, the vertical axis above, $s$, is primarily a
function of temperature. The horizontal axis is tuned by
$\overline{\eta}$, which is mainly a measure of carrier
concentration away from the commensurate value appropriate to the
$p\times p$ ordered state. There are three stable mean field
phases separated by a first order transition (thick line), a
second order transition (solid line), and a second-order critical
point at $s=\overline{\eta}=0$. The phases A, B, C, are described
in the text. For phases A and B we show a schematic of the density
modulation $\delta \rho ({\bf r})$ for the special case $p=4$,
with the points representing the underlying lattice. Upon
including fluctuations, the transition between phases A and B can
become second-order; in this situation the phase adjacent to phase
A is a new phase B$_S$, the incommensurate striped phase. The
phase B$_S$ preempts a portion (or all) of phase B. See
Fig~\ref{figbs} for the fluctuation-corrected phase diagram of
phases A and B.} \label{figmft}
\end{figure}

We will begin in Section~\ref{sec:mft} by defining the order
parameters of the commensurate $p \times p$ solid on the square
lattice. Symmetry considerations then allow us to obtain a
phenomenological free energy density controlling thermal
fluctuations of the order parameter, and a corresponding mean
field phase diagram shown in Fig.~\ref{figmft}.
At this stage, this phase diagram contains 3 phases whose
characteristics we summarize below:
\newline
(A) {\bf Commensurate solid:} This is the ground state with
long-range density wave order with period $p$ along both the $x$ and
$y$ axes.
\newline
(B) {\bf Incommensurate, floating solid:} In mean field theory, this
phase has density waves with the same incommensurate period along
both the $x$ and $y$ directions. This order is only quasi-long-range
{\em i.e.\/} density wave correlations decay with a power-law at
long distances.
\newline
(C) {\bf Liquid:} This is the high temperature phase in which all
density wave correlations decay exponentially with distance.

The remainder of the paper will consider fluctuation corrections to
this mean field phase diagram.

In mean field theory, the transition between phases A and B is
first order, with a jump in the period of the density wave from a
commensurate to an incommensurate value. We know from previous
work on anisotropic phases \cite{pt,sue,schulz,haldane} that such
commensurate-incommensurate transitions are driven by the
proliferation of domain walls, and we consider the domain wall
theory for such a transition in Section~\ref{sec:domains}. In our
case there are two sets of domain walls, running predominantly in
the $x$ and $y$ directions, and a key parameter will be the
intersection energy, $f_I$, of these domain walls. If $f_I < 0$,
then the mean field prediction of a first order transition to an
isotropic floating solid is maintained. However, for the case of a
positive intersection energy ($f_I > 0$), we will demonstrate that
there is a second order transition to an anisotropic,
incommensurate ``striped'' state, which we denote B$_S$. The B$_S$
state has commensurate long range order with a period of $p$
lattice along one axis, and incommensurate quasi-long-range order
along the orthogonal axis. Because B$_S$ has long-range order only
along one direction, it may also be labelled a `smectic'
\cite{chaikinlub,kfe}; however the presence of the underlying
lattice makes its fluctuations quite different from a conventional
smectic liquid crystal. The striped solid B$_S$ will replace a
portion of phase B, so that B$_S$ covers at least the entire
second-order phase boundary to phase A (see Fig~\ref{figbs}). The
remainder of phase B could remain an isotropic floating solid, or
it could be anisotropic in its entirety.

Section~\ref{sec:floating} will consider the transitions from the
incommensurate states in phase B to the disordered liquid C. Two
distinct scenarios are possible, depending upon the whether the
transition takes place from an isotropic floating solid B, or from
the anisotropic solid (B$_S$) which has incommensurate order along
only one axis. In either case, the transition is driven by the
unbinding of dislocations \cite{nh,apy}, and the complete theory
for such transitions will be presented.

Finally, Section~\ref{sec:dual} will present the theory of the
direct transition from the commensurate $p \times p$ solid A to
the liquid state C. From Fig.~\ref{figmft} it appears that such a
transition is only possible at special commensurate values of the
density. For the physically relevant case of $p=4$ we will find
that a direct second order transition is possible. It is described
by a self-dual theory with continuously varying exponents, and is
a generalization of the theory found in Ref.~\onlinecite{jkkn} for
the XY model.

\section{Order parameters and mean field theory}
\label{sec:mft}

As noted in Section~\ref{sec:intro}, all phases and transitions
examined here are associated with the order of a generic
`density', which could be any observable invariant under spin
rotations and time reversal. We represent this density by $\delta
\rho ({\bf r})$. At sufficiently low temperatures, in weak
disorder, and for some range of carrier concentration, we assume
that this density prefers to order with a period of $p$ lattice
spacings ($p=4$ is the case of interest for the cuprates). We can
therefore write
\begin{equation}
\delta \rho ({\bf r}) = \mbox{Re} \left[ \Phi_x e^{i {\bf K}_x
\cdot {\bf r}}\right] + \mbox{Re} \left[ \Phi_y e^{i {\bf K}_y
\cdot {\bf r}} \right], \label{defPhi}
\end{equation}
where ${\bf K}_x = (2 \pi /a) (1/p, 0)$, ${\bf K}_y = (2 \pi /a)
(0,1/p)$, and $\Phi_{x,y}$ are complex order parameters which vary
slowly the on the scale of a lattice spacing.

We now want to write down the most general free energy for
$\Phi_{x,y}$ consistent with the symmetries of the underlying
lattice. Among these are $T_{x,y}$ which translate by a lattice
spacing in the $x,y$ directions, and $I_{x,y}$ which reflect the
$x,y$ axes. These operations lead to
\begin{eqnarray}
T_x &:& \Phi_x \rightarrow \Phi_x e^{2 i \pi/p}~~;~~\Phi_y
\rightarrow \Phi_y \nonumber \\
T_y &:& \Phi_x \rightarrow \Phi_x ~~;~~\Phi_y
\rightarrow \Phi_y e^{2 i \pi/p} \nonumber \\
I_x &:& \Phi_x \rightarrow \Phi_x^{\ast}~~;~~\Phi_y
\rightarrow \Phi_y \nonumber \\
I_y &:& \Phi_x \rightarrow \Phi_x~~;~~\Phi_y \rightarrow
\Phi_y^{\ast}. \label{t1}
\end{eqnarray}
We will also assume the symmetry of rotations by 90 degrees, $R$,
under which
\begin{equation}
R : \Phi_x \rightarrow \Phi_y~~;~~\Phi_y \rightarrow
\Phi_x^{\ast}. \label{t2}
\end{equation}
This symmetry is absent in some of the cuprate compounds (most
notably, in YBCO), and it is not difficult to extend our
considerations to include this case.

We can now write down the most general free energy density,
expanded in powers of $\Phi_{x,y}$ and its gradients, consistent
with the symmetries in Eqs.~(\ref{t1}) and (\ref{t2}) (see also
Refs.~\onlinecite{zachar} and~\onlinecite{ying}). This is
\begin{eqnarray}
\mathcal{F}_{\Phi} &=& \int d^2 r \Biggl[ C_1 \left( \left|
\partial_x \Phi_x \right|^2 + \left| \partial_y \Phi_y \right|^2
\right) \nonumber \\
&~&~~~~~~+ C_2 \left( \left|
\partial_y \Phi_x \right|^2 + \left| \partial_x \Phi_y \right|^2
\right) \nonumber \\
&~&~~~~~~+ i \overline{\eta} \left( \Phi_x^{\ast} \partial_x
\Phi_x +
\Phi_y^{\ast} \partial_y \Phi_y \right) \nonumber \\
&~&~~~~~~+ s \left( |\Phi_x |^2 + |\Phi_y |^2 \right) +
\frac{u}{2} \left(|\Phi_x |^2 + |\Phi_y |^2 \right)^2 \nonumber
\\ &+& v |\Phi_x |^2 |\Phi_y |^2 + w \Phi_x^p + \mbox{c.c.} +
w \Phi_y^p + \mbox{c.c.}  \Biggr]. \label{FP}
\end{eqnarray}

Here $s$ is a parameter which we will tune to drive the
transition; it is assumed to contain the primary dependence on
temperature, and will also have some dependence on carrier
density.

The term proportional to $\overline{\eta}$ is allowed by the
symmetries. It indicates that at sufficiently high temperature,
the density correlations are generically {\em incommensurate\/}.
This incommensurability is a consequence of the values of the
domain wall energies between the $p$ commensurate ordered states
\cite{ostlund,huse}: a domain wall between state 1 and state 2
(say) will generally have a different energy than a domain wall
between state 1 and state $p$. In other words, if we represent the
$p$ states along one direction by a $p$ state clock model, then
the interactions of the clock model are `chiral'. \cite{ostlund}

The sign of the parameter $v$ implies a preference for either
isotropic order ($v<0$ prefers both $\langle \Phi_x \rangle$ and
$\langle \Phi_y \rangle$ non-zero), or anisotropic order ($v>0$
prefers only one of $\langle \Phi_x \rangle$ or $\Phi_y$
non-zero). We will assume throughout this paper that $v<0$, so
that $\langle \Phi_x \rangle$ and $\langle \Phi_y \rangle$ are
both non-zero at sufficiently low temperatures. Nevertheless, we
will find that thermal fluctuations can induce an anisotropic
striped state over an intermediate temperature range.

The complex parameter $w$ accounts for the commensurability
lock-in energy. It ensures that at sufficiently low temperatures
(specifically, for $s$ sufficiently negative), the order has a
commensurate period of $p$ lattice spacings. For $v<0$ (assumed
throughout), the ordering will be $p \times p$. The phase of $w$
controls the phase of this ordering (`site-centered' or
`bond-centered'). None of our results will be sensitive to this
phase, and will apply equally to all of them.

\subsection{Mean field theory}
\label{sec:mfta}

Now we present the results of a simple mean field minimization of
$\mathcal{F}_\Phi$, leading to the phase diagram in
Fig.~\ref{figmft}. As described in the introduction
we find three distinct phases.  The low temperature ground state, labelled by
A has long rage density wave order with period $p$ characterized by
\begin{equation}
\Phi_x = \Phi_y \propto e^{i \theta(p)}
\end{equation}
where $\theta(p) = (2n+1)\pi /p$ if the phase is bond-centered and $\theta(p)
= 2n\pi/p$ if the phase is site-centered with $n=0,1,\ldots,p-1$. The
inset in Fig.~\ref{figmft} A shows the locked in solid for the
special case of a bond-centered phase with $p=4$ where the positions of the
underlying atoms are indicated by the light gray circles.  This commensurate phase
has a first order transition described by the line
\begin{equation}
s = -\left(\frac{\sqrt{2u+v-pw}}{\sqrt{2u+v}-\sqrt{2u+v-pw}}\right)
    \frac{\overline{\eta}^2}{4C_1}
\end{equation}
to B which has incommensurate or floating density wave order
characterized by
\begin{equation}
\Phi_x \propto e^{i (\overline{\eta} / 2C_1)x} \quad \mbox{;}
\quad \Phi_y \propto e^{i (\overline{\eta} / 2C_1)y}.
\end{equation}
The magnitude of both $\Phi_x$ and $\Phi_y$ are equal and they have
the same incommensurate period along perpendicular directions. For
the special case of $p=4$ the floating state is shown as an inset in
Fig.~\ref{figmft} B. The floating solid can melt via a second
order phase transition defined by the line
\begin{equation}
s = \frac{\overline{\eta}^2}{4C_1}
\end{equation}
to a fully disordered state C with $\Phi_x=\Phi_y=0$. Finally, there is a
tri-critical point for $s=\overline{\eta}=0$ where the commensurate
solid A can melt directly to the liquid phase C.

\section{Domain wall theory of the commensurate-incommensurate transition}
\label{sec:domains}

In Section~\ref{sec:mfta} we found a mean-field first order
transition between the commensurate $p\times p$ solid A and the
incommensurate, isotropic floating solid B. This section will
examine fluctuations near this transition more carefully. We will
find that the transition can actually be second order under
suitable conditions, and the second-order order transition is to
an anisotropic, striped state with commensurate $p$ period order
along one direction, and incommensurate order along the orthogonal
direction.

For our study of the initial melting of the commensurate ordered
state, we will assume that dislocations can be ignored. The effect
of dislocations will be considered in Section~\ref{sec:floating},
and we will then verify the self-consistency of this assumption.
Instead, the primary actors will be domain walls between the
commensurate states, as is also the case in previous theories
\cite{pt} of the commensurate-incommensurate transition in
anisotropic systems.

In the absence of dislocations, we can focus on globally defined
single-valued angular variables $\theta_{x,y}$ with
\begin{equation}
\Phi_x \propto e^{i \theta_x}~~;~~ \Phi_y \propto e^{i \theta_y}.
\label{deftheta}
\end{equation}
Note that the values of $\theta_{x,y}$ span over all real numbers,
and not just modulo $2 \pi$. However, periodic boundary conditions
need only be satisfied modulo $2 \pi$. In the na\"ive continuum
limit, the free energy for $\theta_{x,y}$ can be expanded in
powers of the local ``strains'' $\nabla_{\bf r} \theta_{x,y}$:
\begin{eqnarray}
\mathcal{F}_\theta &=& \int d^2 r \Biggl[ \frac{K_1}{2} \left[
(\partial_x \theta_x)^2 + (\partial_y \theta_y )^2 \right]
\nonumber \\
&+& \frac{K_2}{2}\left[ (\partial_x \theta_y)^2 + (\partial_y
\theta_x )^2 \right] + K_3 (\partial_x \theta_y)(
\partial_y \theta_x)
\nonumber \\
&+& K_4 (\partial_x \theta_x) (\partial_y \theta_y) - \eta \left[
\partial_x \theta_x + \partial_y \theta_y \right] \nonumber \\
&-& h \left[ \cos ( p \theta_x ) + \cos (p \theta_y) \right] -
\ldots \Biggr]. \label{ftheta}
\end{eqnarray}
All terms above are, in principle, obtained from those in
Eq.~(\ref{FP}), but now we have retained terms up to second order
in spatial gradients. Now, the incommensuration is induced by the
total derivative terms proportional to $\eta$: these are non-zero
because the angular fields can accumulate an integer multiple of
$2 \pi$ even under periodic boundary conditions. Similarly, we do
not have the freedom to integrate by parts, and so combine the
terms proportional to $K_3$ and $K_4$. The commensurability energy
is now imposed by the $p$-fold field $h_p$.

We will begin in Section~\ref{sec:solitons} by describing the
mean-field structure of the domain wall (or `soliton') excitations
of $\mathcal{F}_\theta$. Then, in Section~\ref{sec:wandering} we
will present the theory of the commensurate-incommensurate
transition driven by the proliferation of these domain walls.
Section~\ref{sec:spinwave} will address the nature of density
fluctuations within the incommensurate phases: we will estimate
the elastic constants of these phases using the domain wall theory
of Section~\ref{sec:wandering}.

\subsection{Energetics of domain walls}
\label{sec:solitons}

Consider starting at low temperatures from a fully ordered
two-dimensional crystal. In this state $\theta_x = 2 \pi n/p$,
$\theta_y = 2\pi n' /p$, where $n$, $n'$ are integers, at all
points in space. We are now interested in the deviations from this
perfectly ordered state as measured by the continuum free energy
$\mathcal{F}_\theta$ in Eq.~(\ref{ftheta}).

The simplest deviation is a single line domain wall in which
$\theta_x$ increases by $2 \pi/p$ across the domain wall. For
$\eta > 0$ (which we assume, without loss of generality), such a
domain wall will preferentially run in the $y$ direction {\em
i.e.\/} the domain wall has $\theta_y$ constant, and $\theta_x$ a
function of $x$ only. The $x$ dependence of $\theta_x$ can be
determined by the sine-Gordon saddle point equation
\begin{equation}
K_1 \partial_x^2 \theta_x(x) = ph\sin\left[p\theta_x (x)\right]
\end{equation}
subject to the boundary conditions
\begin{equation}
\theta_x (0) = 0 \quad \mbox{;} \quad \theta_x (L_x) =
\frac{2\pi}{p} \label{thetaBC}
\end{equation}
where $L_x$ is the size of the system in the $x$-direction. The
solution is a soliton with equation
\begin{equation}
\theta_x (x) =
\frac{4}{p}\tan^{-1}\left[e^{p\sqrt{h/K_1}(x-L_x/2)}\right]
\end{equation}
which allows us to determine the value of the domain wall free energy
per unit length (excluding the contribution of the $\eta$ term in
$\mathcal{F}_I$), which we represent by $\epsilon$,
\begin{equation}
\epsilon = \frac{8}{p}\sqrt{K_1 h}.
\end{equation}
Similarly, there is a corresponding domain wall in $\theta_y$,
with identical physical properties.

\begin{figure}
\centering
\includegraphics[width=3in]{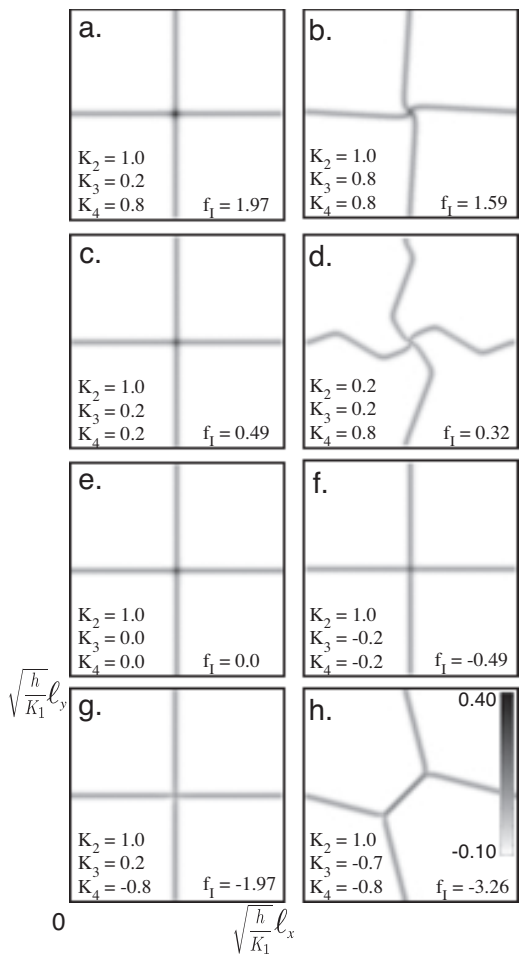}
\caption{Eight possible configurations of the free energy density
for domain wall crossings depending on the numerical values of
$K_2$, $K_3$ and $K_4$ with $p=4$.  The grey scale plots the size of the
local free energy density and all stiffnesses and energies are in units of
$K_1$. Panels a. through h. show decreasing domain wall
intersection energy $f_I$, and for straight walls, the numerically
computed value can be compared with $(\pi^2/4) K_4$.}
\label{figdomain}
\end{figure}

Now we consider the interesting case with domain walls running in
both the $x$ and $y$ directions. These walls will intersect, and
we are interested in the nature of the intersection, and of the
intersection free energy $f_I$.  We can focus in on a single domain wall
crossing by imposing the boundary conditions of Eq.~(\ref{thetaBC}) in both
the $x$ and $y$ directions. For the special case of perfectly straight
domain walls, $\theta_x(x,y) = \theta_x(x)$ and $\theta_y(x,y) =
\theta_y(y)$, the only interaction term in Eq.~(\ref{ftheta}) can be
integrated exactly to give
\begin{equation}
f_I = \left(\frac{2\pi}{p}\right)^2 K_4
\label{eqfi}
\end{equation}
indicating that the sign of the domain wall intersection energy is equal to
the sign of $K_4$.  When the domain walls are not straight, the interaction
energy can be found by numerically minimizing $\mathcal{F}_\theta$ and
evaluating the interaction terms at the ground state field
configurations as seen in Fig.~\ref{figdomain}.
For the following discussion we refer to Fig.~\ref{figdomain} and it
is assumed that all stiffnesses are measured in units of $K_1$.
For positive values of $K_4$ and $K_2=1.0$ (panels a through
c) we observe straight domain walls for $K_3 < K_2$ and only
observe deviation when $K_3 \simeq K_2$.
In panel d, $K_3 \simeq K_2 < K_4$ and we find that the walls
wander quite significantly.  Panels a through d all have $K_4 > 0$
and in agreement with our earlier discussion, $f_I > 0$ in all
four configurations.  Panel e has no interaction terms ($K_3=K_4=0$)
and consequently $f_I = 0$.  For negative values of $K_4$ we
find that the domain wall intersection energy changes to a
negative value.  For $|K_3| < K_2$ the walls remain straight
(panels f and g) but as $K_3 \simeq K_4 \simeq -K_2$ the
intersection energy becomes large and negative and the system attempts
to extend the domain wall overlap over a finite region (panel h).

Although we have only presented eight distinct configurations here,
all of the $K_j$ parameter space was examined.  When the magnitude
of $K_3$ and $K_4$ are small with respect to $K_2$ and $K_1$ we always
find configurations with domain walls crossing at right angles where
the domain wall intersection energy can be calculated using
Eq.~(\ref{eqfi}).

\subsection{Proliferation of domain walls}
\label{sec:wandering}

Now we imagine increasing the parameter $\eta$ so that the total
free energy per unit length of a domain wall is eventually
negative. In such a situation we expect a proliferation of domain
walls, leading to the appearance of a floating solid with
incommensurate density correlations. This section will discuss the
theory of such a transition.

Let the incommensurate state have domain walls in $\theta_x$ with
an average spacing $\ell_x$, and domain walls in $\theta_y$ with
an average spacing $\theta_y$. From the energy per unit length of
these domain walls, and their intersection energy, computed in
Section~\ref{sec:solitons}, there is a clearly a contribution to
the free energy per unit area, $\mathcal{F}_d$, given by (see
Fig.~\ref{figdomain})
\begin{equation}
\mathcal{F}_d^{(1)} (\ell_x , \ell_y) = \left( \epsilon - \frac{2
\pi \eta}{p} \right) \left( \frac{1}{\ell_x} + \frac{1}{\ell_y}
\right) + \frac{f_I}{\ell_x \ell_y}. \label{fd1}
\end{equation}

\begin{figure}
\centering
\includegraphics[width=3in,height=3in]{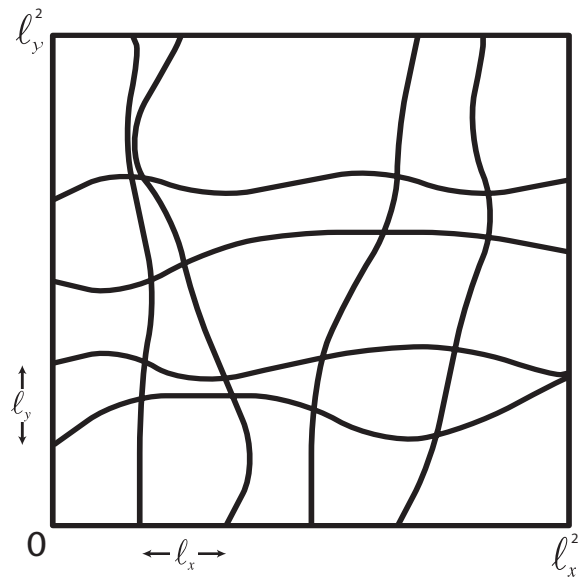}
\caption{A net of wandering domain walls constructed using random walkers on
a lattice with both hard-core repulsion and restricted phase space. If
the domain walls are separated on average by a
distance $\ell_x$ in the $x$ direction and $\ell_y$ in the $y$
direction, then we observe collisions between domain walls running in the
same direction with separation on the order of $\ell_x^2$ or $\ell_y^2$
($\ell_x = \ell_y$ here).}
\label{figdwnet}
\end{figure}

However, in addition to the simple energetic contributions in
Eq.~(\ref{fd1}), we also have to consider the entropic
contribution of the wandering of the domain walls (Fig.~\ref{figdwnet}).
This can be computed using the elegant free fermion mapping of
Pokrovsky and Talapov \cite{pt}, which applies here (essentially unchanged)
separately to the domain walls in each direction. Briefly, the
argument runs as follows. Let $u_x (y)$ represent the $x$
co-ordinate of a domain wall in $\theta_x$. Any $y$ dependence in
$u_x$ increases the total length of the domain wall, and so leads
to a free energy cost
\begin{equation}
\mathcal{F}_w = \frac{1}{2} \epsilon \int dy\, \left(
\frac{du_x}{dy} \right)^2. \label{fwall}
\end{equation}
Now the
partition function at a temperature $T$ of such domain walls can
be mapped onto that for free fermions with a density $1/\ell_x$
and mass $\epsilon/T$. From the ground state energy of such a free
fermion system, we then obtain the contribution of the wandering
of the domain walls to the free energy density
\begin{equation}
\mathcal{F}_d^{(2)} (\ell_x, \ell_y) = \frac{\pi^2 T^2}{6
\epsilon} \left( \frac{1}{\ell_x^3} + \frac{1}{\ell_y^3} \right).
\end{equation}

\begin{figure}
\centering
\includegraphics[width=3in]{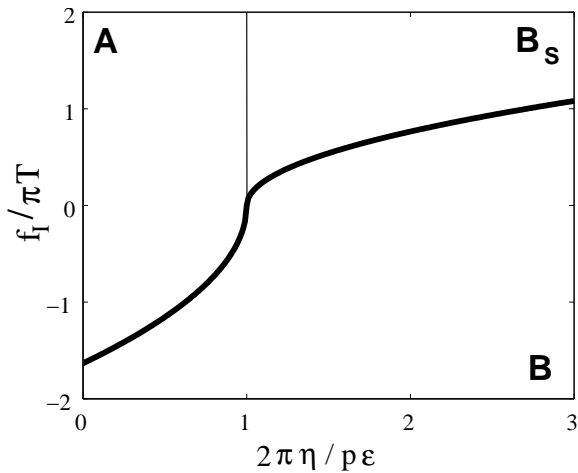}
\caption{Phase diagram of the commensurate-incommensurate
transition obtained by the minimization of Eq.~(\ref{fd}) over the
values of the mean domain wall spacing $\ell_x$ and $\ell_y$.}
\label{figbs}
\end{figure}

We are now faced with the simple problem of minimizing the free
energy density
\begin{equation}
\mathcal{F}_d (\ell_x, \ell_y) = \mathcal{F}_d^{(1)} (\ell_x,
\ell_y) + \mathcal{F}_d^{(2)} (\ell_x, \ell_y) \label{fd}
\end{equation}
as a function of $\ell_x$ and $\ell_y$ to determine the nature of
the commensurate-incommensurate transition. This simple
calculation turns out to have some interesting structure which we
will now describe. The results of such a minimization are
summarized in Fig~\ref{figbs}.
The nature of the phases appearing here depends upon the sign of
the domain wall intersection energy $f_I$.

For negative $f_I$ corresponding to $K_4 < 0$ in
$\mathcal{F}_\theta$, as we increase the value of $\eta$ the mean
field prediction is preserved, and the ground state with long
range charge density wave order melts via a first order transition
to an isotropic incommensurate floating solid which we labelled as
B. This transition occurs when
\begin{equation}
\eta_{A B} = \frac{p\epsilon}{2\pi}\left[1 - \frac{3}{8}
        \left(\frac{f_I}{\pi T}\right)^2 \right].
\end{equation}
That is, for $\eta < \eta_{A B}$ the system is commensurate with
$1/\ell_x = 1/\ell_y = 0$ but for $\eta > \eta_{A B}$ there is a
jump to the isotropic floating state B with domain wall densities
\begin{equation}
\frac{1}{\ell_x} = \frac{1}{\ell_y} = -\frac{f_I \epsilon}{\pi^2
T^2} \label{lxB} + \sqrt{ \left( \frac{ f_I \epsilon}{\pi^2 T^2}
\right)^2 - \frac{2 \epsilon (\epsilon - 2 \pi \eta/p)}{\pi^2
T^2}}.
\end{equation}

If the domain wall intersection energy is positive, implying that
collisions are disfavored ($f_I > 0$, $K_4 > 0$), then as we
increase the value of $\eta$ from zero the state remains
commensurate with long range order and no domain walls ($1/\ell_x
= 1/\ell_y = 0$) until
\begin{equation}
\eta_{A B_S} = \frac{p\epsilon}{2\pi}
\end{equation}
where there is a second order phase transition to a state with
(say)
\begin{equation}
\frac{1}{\ell_y} = 0 \quad \mbox{;} \quad
\frac{1}{\ell_x} = \frac{1}{\pi T} \sqrt{2\epsilon\left( \frac{2\pi
\eta}{p} - \epsilon \right)}
\label{lxBI}
\end{equation}
which has domain walls running in \emph{either} the $x$ or $y$
direction but not \emph{both}.  This anisotropic floating solid or
incommensurate striped state which we label B$_S$ has long range
charge density wave order with period $p$ in one direction and
quasi long range order with power law decay in the perpendicular
direction. Near the second-order transition between A and B$_S$,
insertion of Eq.~(\ref{lxBI}) into Eq.~(\ref{fd}) shows that
$\mathcal{F}_d \sim - (\epsilon - 2\pi \eta/p)^{3/2}$.

As $\eta$ is increased further for $f_I >0$, the phase B$_S$
persists until
\begin{equation}
\eta_{B_S B} = \frac{p\epsilon}{2\pi}\left[1 +
        \left(1.70968\ldots\right)\left(\frac{f_I}{\pi T}\right)^2 \right]
\end{equation}
where there is a first order transition to the isotropic
incommensurate floating solid B with $\ell_x$ and $\ell_y$ still
given by Eq.~(\ref{lxB}). In fact, the solution in Eq.~(\ref{lxB})
is a local minimum of the free energy for all $\eta > p
\epsilon/(2 \pi)$. However, its free energy behaves like
$\mathcal{F}_d \sim - (\epsilon - 2\pi \eta/p)^{2}$ upon
approaching phase A. Close enough to phase A, this free energy is
always larger than the free energy for phase B$_S$.

Therefore, if the intersection energy for domain walls is positive,
the melting of the commensurate solid A always occurs via a
second-order transition to the striped anisotropic state B$_S$.

\subsection{Renormalization of elastic constants in incommensurate phases}
\label{sec:spinwave}

Once the domain walls have proliferated in phases B and B$_S$, the
density correlations become incommensurate and appear at
wavevectors which are shifted from the commensurate values in
Eq.~(\ref{defPhi}). We are interested here in the long wavelength
``spinwave'' density fluctuations about this incommensurate state.

First, let us consider the phase B. Rather than defining the
spin-wave variables about the commensurate state as in
Eq.~(\ref{deftheta}), we now need to look at fluctuations above an
incommensurate ordered state. So now we write
\begin{equation}
\Phi_x \propto e^{i 2 \pi x/(p \ell_x)} e^{i \theta_x}~~;~~ \Phi_y
\propto e^{i 2 \pi y/(p \ell_y)} e^{i \theta_y}, \label{deftheta2}
\end{equation}
because the incommensurate ordering wavevectors are $ (2 \pi /a)
(1/p+a/\ell_x, 0)$ and $(2 \pi /a) (0,1/p+a/\ell_y)$. We are
interested in the effective action for these $\theta_{x,y}$ on a
coarse-grained scale much larger than $\ell_{x,y}$. This action
can be deduced by the same symmetry arguments made in
Section~\ref{sec:mft}. It is not difficult to see that this action
in the floating phase B has the same structure as
$\mathcal{F}_\theta$ in Eq.~(\ref{ftheta}), except that now
$\eta=h=0$:
\begin{eqnarray}
\mathcal{F}_{{\rm B}} &=& \int d^2 r \Biggl[ \frac{K_1}{2} \left[
(\partial_x \theta_x)^2 + (\partial_y \theta_y )^2 \right]
\nonumber \\
&+& \frac{K_2}{2}\left[ (\partial_x \theta_y)^2 + (\partial_y
\theta_x )^2 \right] + K_3 (\partial_x \theta_y)(
\partial_y \theta_x)
\nonumber \\
&+& K_4 (\partial_x \theta_x) (\partial_y \theta_y) \Biggr]
\label{ftheta1}
\end{eqnarray}

For the phase B$_S$, the $\theta_y$ variables (say) are locked at
their commensurate value, and so only the $\theta_x$ variables
will contribute to the spin-wave theory:
\begin{eqnarray}
\mathcal{F}_{{\rm B}_S} &=& \int d^2 r \Biggl[ \frac{K_1}{2}
 (\partial_x \theta_x)^2+ \frac{K_2}{2} (\partial_y
\theta_x )^2  \Biggr] \label{ftheta2}
\end{eqnarray}

An important point is that the stiffnesses $K_{1-4}$ in
Eqs.~(\ref{ftheta1}) and (\ref{ftheta2}) be strongly renormalized
from the bare values in Eq.~(\ref{ftheta}). Following the analysis
of Ref.~\onlinecite{sue}, we will now estimate their renormalized
values $K_{1-4}^R$ in terms of the parameters appearing in the domain
wall free energies in Section~\ref{sec:wandering} (the superscript
$R$ will be used for clarity only in this subsection).

First, imagine that we impose a uniform compressional strain
$\partial_x \theta_x$ on the floating solid. On the average, this
will cause the domain walls to move closer to each other, and
change the value of $\ell_x$ to $\ell_x - \delta \ell_x$. Because
$\theta_x$ changes by $2\pi/p$ across each domain wall, we
conclude that $\partial_x \theta_x = (2 \pi/p) \delta \ell_x
/\ell_x^2$. The change in $\ell_x$ will cause a change in free
energy that can be computed from Eq.~(\ref{fd}), and so we
conclude
\begin{equation}
K_1^R = \frac{p^2 \ell_x^4}{4 \pi^2} \frac{\partial^2
\mathcal{F}_d}{\partial \ell_x^2},
\end{equation}
where the derivative has to be computed at the equilibrium values
of $\ell_x$, $\ell_y$ which minimize Eq.~(\ref{fd}). This
determines
\begin{equation}
K_1^R = \frac{p^2 T^2}{4 \epsilon \ell_x},
\end{equation}
in both phases B and B$_S$. Similarly, we can apply a combined
compressional strain in $\theta_x$ and $\theta_y$ and conclude
\begin{equation}
K_4^R = \frac{p^2 \ell_x^2 \ell_y^2}{4 \pi^2} \frac{\partial^2
\mathcal{F}_d}{\partial \ell_x \partial \ell_y},
\end{equation}
This is non-zero only in the floating phase B, and we obtain
\begin{equation}
K_4^R =\frac{p^2 f_I}{4\pi^2}~~,~~\mbox{phase B only}.
\end{equation}

To determine $K_2^R$, apply a small uniform shear strain to
$\theta_x$ by inducing a non-zero $\partial_y \theta_x$. This will
move the domain walls in $\theta_x$ such that the position of each
domain wall obeys $\partial_y u_x = (p \ell_x/(2 \pi)) \partial_y
\theta_x$. Inserting this into Eq.~(\ref{fwall}), we obtain
\begin{equation}
K_2^R = \frac{\epsilon}{\ell_x} \left( \frac{ p \ell_x}{2 \pi}
\right)^2 = \frac{ \epsilon p^2 \ell_x}{4 \pi^2},
\end{equation}
in both phases B and B$_S$.

Finally, to determine $K_3^R$ we need to apply shear strains to
both $\theta_x$ and $\theta_y$. Neither of them causes a net
change in the density of domain walls, or in the number of
intersections between the domain walls. Consequently there is no
change to the free energy beyond that already accounted for by
$K_2^R$, and hence
\begin{equation}
K_3^R = 0,
\end{equation}
in both phases B and B$_S$.

\section{Dislocation mediated melting of floating solids}
\label{sec:floating}

We will now consider the transition from the isotropic floating
solid B and the striped floating solid B$_S$ to the disordered
liquid phase C. These transitions are driven by the unbinding of
dislocations.

With the parameters of the ``spin-wave'' theory at hand in
Section~\ref{sec:spinwave}, we can address the energetics of the
the dislocations. We will derive the effective action for the
dislocations in Section~\ref{sec:disc} and then obtain the
renormalization group (RG) flow equations in
Section~\ref{sec:discrg}.

\subsection{Dislocation interactions}
\label{sec:disc}

For the most part, this section will be restricted to a discussion
of dislocations in phase B. The simpler case of the anisotropic
phase B$_S$ is easily obtained by only including those terms
arising from the fluctuations of $\theta_x$.

Dislocations are simply `vortices' in the angular fields
$\theta_{x,y}$ under which
\begin{eqnarray}
\oint d{\bf r} \cdot \nabla_{\bf r} \theta_x &=& 2 \pi m_x ({\bf
r}_v) \nonumber \\
\oint d{\bf r} \cdot \nabla_{\bf r} \theta_y &=& 2 \pi m_y ({\bf
r}_v) \label{contour}
\end{eqnarray}
where $m_{x,y}$ are integers at the vortex ({\em i.e.\/}
dislocation) site ${\bf r}_v$, and the integral is over a contour
that encloses ${\bf r}_v$. Each dislocation is therefore
characterized by a doublet of integers $(m_x, m_y)$.

To compute the interactions between these vortices, it is useful
to define continuum vortex densities by
\begin{equation}
m_i ({\bf r}) = \sum_v m ({\bf r}_v) \delta ({\bf r} - {\bf r}_v )
\end{equation}
where $i=x,y$. Then, after transforming to momentum (${\bf k}$)
space, the relation Eq.~(\ref{contour}) can be written simply as
\begin{equation}
k_i \theta_j = k_i \vartheta_j + \frac{2 \pi}{k^2}
\epsilon_{i\ell} k_{\ell} m_j \label{vartheta}
\end{equation}
where $\epsilon$ is the antisymmetric tensor, and $\vartheta_j$ is
an arbitrary smooth angular field which has no vortices. We now
insert Eq.~(\ref{vartheta}) into the free energy $\mathcal{F}_{\rm
B}$ in Eq.~(\ref{ftheta1}), and minimize with respect to
$\vartheta_j$. The result for the total free energy of the
vortices is then
\begin{eqnarray}
\mathcal{F}_v &=& \int \frac{d^2 k}{4 \pi^2} \frac{2 \pi^2}{D(k_x,
k_y)} \Biggl[ |m_x ({\bf k}) |^2 \left( k_x^2 K_1 (K_2^2 - K_3^2)
\right. \nonumber \\ &+& \left. k_y^2 K_2 (K_1^2 - K_4^2) \right)
+ |m_y ({\bf k}) |^2 \left( k_y^2 K_1 (K_2^2 - K_3^2) \right.
\nonumber \\ &+& \left. k_x^2 K_2 (K_1^2 - K_4^2) \right) + m_x
({\bf k}) m_y (- {\bf k}) k_x k_y \left(K_1^2 K_3 \right.
\nonumber \\ &+& \left. K_2^2 K_4 - K_3 K_4 (K_3 + K_4 ) \right)
\Biggr] \label{fv}
\end{eqnarray}
where
\begin{eqnarray}
&& D (k_x, k_y) \equiv K_1 K_2 (k_x^2 + k_y^2)^2 \nonumber \\
&&~~~~~+\ k_x^2 k_y^2 \left( (K_1 - K_2)^2 - (K_3 + K_4)^2 \right).
\end{eqnarray}
Note that the values of $K_{1-4}^R$ from
Section~\ref{sec:spinwave} are to be inserted into the expressions
above; here, and henceforth, the superscript $R$ has been dropped.
The interaction between the vortices in now determined by
transforming Eq.~(\ref{fv}) back to real space. This takes the
form
\begin{eqnarray}
\mathcal{F}_v &=& E_c \sum_v \left[ m_x^2 ({\bf r}_v ) + m_y^2
({\bf r}_v ) \right] \nonumber \\
&+& \sum_{v < v'} \Biggl[ m_x ({\bf r}_v) m_x ({\bf r}_{v'})
V\left(|{\bf r}_v - {\bf r}_{v'}|, \phi({\bf r}_v - {\bf
r}_{v'})\right) \nonumber \\
&+&  m_y ({\bf r}_v) m_y ({\bf r}_{v'}) V\left(|{\bf r}_v - {\bf
r}_{v'}|, \phi({\bf r}_v - {\bf r}_{v'})+\pi/2\right)
\nonumber \\
&+&  m_x ({\bf r}_v) m_y ({\bf r}_{v'}) W\left(|{\bf r}_v - {\bf
r}_{v'}|, \phi({\bf r}_v - {\bf r}_{v'})\right) \Biggr]
\end{eqnarray}
where $\phi ({\bf r}) = \arctan(y/x)$ is the angle the vector
${\bf r}$ makes with the $x$ axis. We will not need the explicit
form of the interaction $W$ in our subsequent analysis, and so we
will not specify it explicitly; the interaction $V$ is given by
\begin{eqnarray}
 && V(r, \phi) = \int_0^{\infty} \frac{e^{-ka} dk}{k} \int_0^{2 \pi} d \varphi \left[ e^{i k r
 \cos(\phi - \varphi)} - 1 \right] \nonumber \\
 &&\times \frac{ K_1 (K_2^2 - K_3^2) \cos^2 \varphi +
K_2 (K_1^2 - K_4^2) \sin^2 \varphi}{K_1 K_2 +  \left( (K_1 -
K_2)^2 - (K_3 + K_4)^2 \right) \sin^2 \varphi \cos^2 \varphi }
\nonumber \\
&=& \frac{p^2 T^2}{4\pi^2}  \int_0^{2 \pi} d \varphi
\int_0^{\infty} \frac{e^{-ka} dk}{k} \left[ \cos( k r
 \cos(\phi - \varphi)) - 1 \right] \nonumber \\
 &&~~~~~~~~~~~~~~~~~~~~~~~~~~~~~~~\times \Lambda (\widetilde{K}_i, \varphi)
\nonumber \\
 &\equiv& - \frac{p^2 T^2}{2\pi} \widetilde{L}_0 \ln (r/a) + \widetilde{V} (\phi)
 \label{vrphi}
\end{eqnarray}
where we have inserted a soft cutoff using the lattice spacing
$a$, and
\begin{eqnarray}
&& \!\!\!\!\!\! \Lambda (\widetilde{K}_i, \varphi) \label{deflambda} \\
\!&&= \frac{ \widetilde{K}_2 \cos^2 \varphi + \widetilde{K}_1
\sin^2 \varphi}{\widetilde{K}_1 \widetilde{K}_2 + \left(
(\widetilde{K}_1 - \widetilde{K}_2)^2 - (\widetilde{K}_3 +
\widetilde{K}_4)^2 \right) \sin^2 \varphi \cos^2 \varphi }
\nonumber
\end{eqnarray}
is a function of the new couplings $\widetilde{K}_{1-4}$ defined
by
\begin{eqnarray}
\widetilde{K}_1 = \frac{p^2 T^2}{4\pi^2} \frac{K_2}{K_2^2 - K_3^2}
~~&;&~~ \widetilde{K}_2 = \frac{p^2 T^2}{4\pi^2} \frac{K_1}{K_1^2
- K_4^2}
\nonumber \\
\widetilde{K}_3 = \frac{p^2 T^2}{4\pi^2} \frac{K_4}{K_1^2 - K_4^2}
~~&;&~~ \widetilde{K}_4 = \frac{p^2 T^2}{4\pi^2} \frac{K_3}{K_2^2
-
K_3^2}. \nonumber \\
\label{kdual}
\end{eqnarray}
At this point these definitions of the $\widetilde{K}_{1-4}$ may
be viewed as arbitrary variables, but we will see later in
Section~\ref{sec:dual} and Appendix~\ref{app:dual} that these are
the couplings that appear in a self-dual mapping of the theory
$\mathcal{F}_\theta$. Also, from Appendix~\ref{app:dual}, note
that the $\widetilde{K}_{1-4}$ couplings appear upon taking the
inverse of the matrix of couplings between the strains in
$\mathcal{F}_{\rm B}$; consequently, the inverse expressions for
the $K_{1-4}$ in terms of the $\widetilde{K}_{1-4}$ have exactly
the same structure as in Eq.~(\ref{kdual}). The parameters in the
last line of Eq.~(\ref{vrphi}) are given by
\begin{eqnarray}
\widetilde{L}_0  &=& \int_0^{2 \pi} \frac{ d \varphi}{2 \pi} \Lambda
(\widetilde{K}_i, \varphi)  \label{lkd} \\
&=& \frac{\widetilde{K}_1+\widetilde{K}_2}{\sqrt{\widetilde{K}_1 \widetilde{K}_2
\left[(\widetilde{K}_1+\widetilde{K}_2)^2 -(\widetilde{K}_3 +
\widetilde{K}_4)^2\right]}} \nonumber \\
\widetilde{V} (\phi) &=& -\frac{p^2 T^2}{4\pi^2} \int_0^{2 \pi} d
\varphi \ln \left(|\cos(\phi-\varphi)| \right) \Lambda
(\widetilde{K}_i, \varphi) \nonumber
\end{eqnarray}
where the calculation of $\widetilde{L}_0$ is described in
Appendix~\ref{app:L0}. For $\widetilde{K}_3 = \widetilde{K}_4=0$,
$\widetilde{V} (\phi)$ can be evaluated in closed form:
\begin{eqnarray}
\widetilde{V} (\phi) &=& \frac{p^2 T^2}{4 \pi (\widetilde{K}_1
\widetilde{K}_2)^{1/2}} \ln \left(
\frac{\left(\widetilde{K}_1^{1/2} +
\widetilde{K}_2^{1/2}\right)^2}{\widetilde{K}_1 \sin^2 \phi +
\widetilde{K}_2 \cos^2 \phi} \right). \nonumber \\   \label{wl0}
\end{eqnarray}

\subsection{Renormalization group flows}
\label{sec:discrg}

With the knowledge of the interactions between the dislocations,
the renormalization group equations can be derived by the methods
already described in Refs.~\onlinecite{nh} and~\onlinecite{apy}.
We introduce a vortex fugacity, $y = e^{-E_c/T}$ and examine the
effect of integrating out pairs of dislocations in an expansion in
powers of $y$. A simple and standard analysis shows that the flow
equation for the fugacity is
\begin{equation}
\frac{dy}{d \ell} = \left(2 - \frac{p^2 T}{4\pi} \widetilde{L}_0
\right) y. \label{rgy}
\end{equation}

The renormalization of the $K_i$ from the vortices can be computed
by the method described in Appendix B of Ref.~\onlinecite{apy}. We
compute the renormalization of the elastic constants, $K_i$ by
determining the contribution of the vortices to a two-point
correlation of the strains. Such a procedure leads naturally to
flow equations for the `inverse' or `dual' $\widetilde{K}_i$
couplings, and we obtain
\begin{eqnarray}
\frac{d \widetilde{K}_1}{d\ell} &=& p^2
y^2 T\int_0^{2 \pi} d \phi \cos^2 \phi e^{{\widetilde{V}} (\phi)/T} \nonumber \\
\frac{d \widetilde{K}_2}{d\ell} &=& p^2
y^2 T\int_0^{2 \pi} d \phi \sin^2 \phi e^{\widetilde{V} (\phi)/T} \nonumber \\
\frac{d \widetilde{K}_3}{d\ell} &=& 0\nonumber \\
\frac{d \widetilde{K}_4}{d\ell} &=& 0 \label{rgd}
\end{eqnarray}
We can convert these into equations for the $K_i$ and obtain
\begin{eqnarray}
\frac{d {K}_1}{d\ell} &=& - \frac{4\pi^2 y^2}{T} (K_1^2 + K_4^2 )
\int_0^{2 \pi} d \phi  \sin^2 \phi
e^{\widetilde{V} (\phi)/T }  \nonumber \\
\frac{d {K}_2}{d\ell} &=& - \frac{4\pi^2 y^2}{T} (K_2^2 + K_3^2 )
\int_0^{2 \pi} d \phi  \cos^2 \phi
e^{\widetilde{V} (\phi) /T}  \nonumber \\
\frac{d {K}_3}{d\ell} &=& -\frac{8\pi^2 y^2}{T} K_2 K_3 \int_0^{2 \pi}
d \phi \cos^2 \phi e^{\widetilde{V} (\phi)/T} \nonumber \\
\frac{d {K}_4}{d\ell} &=& -\frac{8\pi^2 y^2}{T} K_1 K_4 \int_0^{2
\pi} d \phi \sin^2 \phi e^{\widetilde{V} (\phi)/T} \label{rgya}
\end{eqnarray}

While complex in appearance, the flow equations Eq.~(\ref{rgy})
and (\ref{rgya}) predict a melting transition of the floating
phase B into the liquid phase C which is in a universality class
closely related to that of the Kosterlitz-Thouless (KT)
transition. The phase B is stable provided $\widetilde{L}_0 > 8
\pi/(p^2 T)$. Evaluating $\widetilde{L}_0$ at the values of the
$K_i^R$ determined in Section~\ref{sec:spinwave}, we find that
phase B is stable everywhere for $p=4$ towards an infinitesimal
vortex fugacity for the elastic constants in the domain wall
theory; this justifies the neglect of dislocations in our study of
domain wall proliferation in Section~\ref{sec:domains}. The
present equations Eq.~(\ref{rgya}) show how phase B will
ultimately become unstable to a vortex unbinding transition once
the vortex fugacity becomes larger. The flow equations in
Eq.~(\ref{rgya}) imply singularities in the elastic constants just
before the melting transition which can be computed as in
Ref.~\onlinecite{ssmelt}. Some of the universal amplitude ratios
here will be different from the KT transition, but all exponents
and critical singularities will be as in the KT transition.

The flow equations for the melting of the anisotropic phase B$_S$
can be easily obtained from Eq.~(\ref{rgya}) simply by setting
$K_3=K_4=0$. In this case, the values of the elastic constants in
Section~\ref{sec:spinwave} and Eq.~(\ref{lkd}) imply that
\begin{equation} \widetilde{L}_0 =
\frac{4 \pi^2}{p^2 T^2} \left(K_1^R K_2^R \right)^{1/2} =
\frac{\pi}{T}.
\end{equation}
The equation for the vortex fugacity Eq.~(\ref{rgy}) implies then
that the vortex fugacity flows to zero as long as\cite{sue} $p^2 >
8$, which is certainly the case for $p>3$. Hence the floating
phase B$_S$ is also initially stable to dislocation unbinding. At
higher temperatures, there can be a dislocation unbinding
transition in the Kosterlitz-Thouless universality class. Note
that in this sequence of transitions, even after such a transition
has occurred from the phase B$_S$, order has only been lost in $x$
direction, and commensurate long-range order remains in
$\theta_y$. So the system is still strongly ``striped'' and
anisotropic. Full isotropy will be restored only after a second
set of similar transitions in $\theta_y$, the first to floating
incommensurate order in $\theta_y$, and then a dislocation
unbinding transition to the liquid phase C.

\section{Direct melting of the commensurate solid}
\label{sec:dual}

As we noted in Section~\ref{sec:intro} and Fig~\ref{figmft}, it is
possible that the low temperature $4\times 4$ commensurate solid
can melt directly into the high temperature liquid state. In the
theories $\mathcal{F}_\Phi$ in Eq.~(\ref{FP}) and
$\mathcal{F}_\theta$ in Eq.~(\ref{ftheta}), such a transition
appears possible at commensurate densities at which
$\overline{\eta} = \eta =0$. So this section will neglect the
influence of $\eta$ and perform a full renormalization group
analysis of the theory $\mathcal{F}_\theta$, including both
dislocations and the $p$-fold lock-in field $h$.

A key property which aids the analysis is a self-duality of the
action in which the dislocation fugacity, $y$, and the lock-in
field $h$ are interchanged. The origin of this self-duality is
similar to that discussed in Ref.~\onlinecite{jkkn}, and we
present a brief derivation in Appendix~\ref{app:dual}. From this
analysis we obtain the the partition function
$\mathcal{F}_\theta$, including dislocations, obeys
\begin{equation}
\mathcal{Z}_\theta (K_1, K_2, K_3, K_4, h, y) = \mathcal{Z}_\theta
(\widetilde{K}_1, \widetilde{K}_2, \widetilde{K}_3,
\widetilde{K}_4, y, h), \label{dual}
\end{equation}
where the couplings $\widetilde{K}_i$ were defined in
Eq.~(\ref{kdual}).

Aided by this duality, we can now immediately deduce the full flow
equations for all the elastic constants $K_i$, the vortex fugacity
$y$ and the field $h$ from the results in
Section~\ref{sec:discrg}. As always, these equations are valid for
small $y$ and $h$ and are
\begin{eqnarray}
\frac{dh}{d \ell} &=& (2 -  \frac{p^2 T}{4\pi} {L}_0 ) h \nonumber \\
\frac{dy}{d \ell} &=& (2 -  \frac{p^2 T}{4\pi} \widetilde{L}_0 ) y
\nonumber \\
\frac{d {K}_1}{d\ell} &=& \int_0^{2 \pi} d \phi \Biggl[ p^2 h^2 T
\cos^2 \phi e^{{V} (\phi)/T} \nonumber \\
&~&~~~~~~~~~~~~- \frac{4\pi^2 y^2}{T} (K_1^2 + K_4^2) \sin^2 \phi
e^{\widetilde{V} (\phi)/T } \Biggr] \nonumber \\
\frac{d {K}_2}{d\ell} &=& \int_0^{2 \pi} d \phi \Biggl[ p^2 h^2 T
\sin^2 \phi e^{{V} (\phi)/T} \nonumber \\
&~&~~~~~~~~~~~~- \frac{4\pi^2 y^2}{T} (K_2^2 + K_3^2) \cos^2 \phi
e^{\widetilde{V} (\phi)/T } \Biggr] \nonumber \\
\frac{d {K}_3}{d\ell} &=& -\frac{8 \pi^2
y^2}{T} K_2 K_3 \int_0^{2 \pi} d \phi \cos^2 \phi e^{\widetilde{V} (\phi)/T} \nonumber \\
\frac{d {K}_4}{d\ell} &=& -\frac{8 \pi^2 y^2}{T} K_1 K_4 \int_0^{2
\pi} d \phi \sin^2 \phi e^{\widetilde{V} (\phi)/T} \label{rgall}
\end{eqnarray}
Here $L_0$ and $V(\phi)$ are defined as in Eq.~(\ref{lkd}), but
with direct couplings:
\begin{eqnarray}
{L}_0  &=& \int_0^{2 \pi} \frac{ d \phi}{2 \pi} \Lambda ({K}_i, \phi)  \\
{V} (\phi) &=& -\frac{p^2 T^2}{4\pi^2} \int_0^{2 \pi} d \varphi
\ln \left(|\cos(\phi-\varphi)| \right) \Lambda ({K}_i, \varphi)
\nonumber \label{lk}
\end{eqnarray}

We will confine our analysis of Eq.~(\ref{rgall}) to the
physically important case of $p=4$. For this case, we first
searched for an intermediate phase with power-law correlations:
such a phase would obtain if there was a set of values of
$K_{1-4}$ for which {\em both\/} $y$ and $h$ flowed to zero. As
shown in Appendix~\ref{app:L0}, there is no such regime of
parameters.

However, the flow equations in Eq.~(\ref{rgall}) do predict a
direct second-order transition between the $4 \times 4$
commensurate solid A and isotropic liquid C. This transition is
described by a manifold of fixed points. In the 6-dimensional
space of couplings, there is 2-dimensional manifold of fixed
points specified by
\begin{equation}
y=h~~;~~\frac{K_1 K_2}{T^2} = \frac{4}{\pi^2}~~;~~K_3=K_4=0.
\end{equation}
In order to describe the flows near this manifold, it is
convenient to make a change of variables from $y$, $h$, $K_1$, and
$K_2$ to
\begin{eqnarray}
\alpha = y-h~~&,&~~\beta=y+h \nonumber \\
K = \frac{\sqrt{K_1 K_2}}{T}-\frac{2}{\pi}~~&,&~~ \lambda =
\sqrt{\frac{K_1}{K_2}}.
\end{eqnarray}
In these variables, the fixed point manifold is described by
$\alpha=K=K_3=K_4=0$, while the values of $\lambda$ and $\beta$
are arbitrary. All physical properties, including the exponents at
the second-order critical point, will depend upon the bare values
of $\lambda$ and $\beta$. We expanded Eqs.~(\ref{rgall}) to linear
order in $\alpha$, $K$, $K_3$, and $K_4$, and after
diagonalization of the flow equations, obtained the following
renormalization group eigenvalues in this 4-dimensional subspace:
\begin{eqnarray}
&& 2 \pi^2 \beta \left(\sqrt{\lambda} + 1/\sqrt{\lambda} \right)^4
\nonumber \\
&&~~~~~~~~\times \left( 1 \pm \left( 1+ \frac{4}{\pi^2 \beta
\left(\sqrt{\lambda} +
1/\sqrt{\lambda} \right)^4}\right)^{1/2} \right),\nonumber \\
&&~~~~- 4\pi^2 \beta^2 \left(\sqrt{\lambda} + 1/\sqrt{\lambda}
\right)^4,
\end{eqnarray}
where the last eigenvalue is doubly degenerate. It is evident that
for all $\lambda$, $\beta$ there are 3 negative eigenvalues and 1
positive eigenvalue. This flow therefore describes a conventional
second-order transition, with the correlation length exponent
$\nu$ equal to the inverse of the positive eigenvalue.

\section{Conclusions}
\label{sec:conc}

We have shown that thermal fluctuations on $4 \times 4$ ordered
state lead generically to a rather rich phase diagram as a
function of temperature and carrier concentration. This phase
diagram generically must have regions with incommensurate and
anisotropic ordering. Thus it is remains within the realm of
possibility that the underlying physics of density wave ordering
in all the cuprates is the same, and the distinctions between the
experiments are entirely due to their distinct locations in our
phase diagram.

For easy reference, we conclude by listing the various routes the
$4 \times 4$ solid A can melt into the liquid C. The list below is
not exhaustive, and omits certain possibilities involving strong
first order transitions between unrelated phases.
\begin{enumerate}
\item Phase A undergoes a second order Pokrovsky-Talapov \cite{pt}
(PT) transition to the incommensurate striped phase B$_S$, as
described in Section~\ref{sec:wandering}. This is followed by a
first order transition into phase B as also described in
Section~\ref{sec:wandering} and Fig~\ref{figbs}. Then phase B
undergoes a dislocation mediated melting transition into phase C
which is described by Eqs.~(\ref{rgy}) and (\ref{rgya}). This last
transition is in a universality class which is nearly, but not
exactly, KT. \item PT transition from phase A to phase B$_S$ as in
1. Phase B$_S$ then has a KT transition to a striped state which
has long range order with period 4 along one direction, and
exponentially decaying correlations along the other. This striped
set melts into C after a second set of similar transitions: a PT
transition into a state with incommensurate quasi-long range order
in one direction only, and then finally a KT transition in C.
\item First order transition from A to B as described in
Section~\ref{sec:wandering} and Fig~\ref{figbs}. Then, a
transition from B to C as in 2. \item At special commensurate
densities, there is a direct, self-dual, second-order transition
from A to C, described in Section~\ref{sec:dual}, with
continuously varying exponents.
\end{enumerate}

\acknowledgments We thank E.~Fradkin, S.~Kivelson, and T.~Lubensky
for valuable discussions. This research was supported by the
National Science Foundation under grant DMR-0098226, and under
grants DMR-0210790, PHY-9907949 at the Kavli Institute for
Theoretical Physics. S.S. was also supported by the John Simon
Guggenheim Memorial Foundation.

\appendix

\section{Evaluation of $\widetilde{L}_0$}
\label{app:L0}

This appendix will present the derivation of $\widetilde{L}_0$ in
Eq.~(\ref{lkd}) and its application to the flow equations presented in
Section~\ref{sec:dual}.  We begin by writing the integral explicitly as
\begin{eqnarray}
\widetilde{L}_0 &=& \int_{0}^{2\pi} \frac{d \varphi}{2\pi} \label{intL0} \\
&\times&\!\! \frac{ \widetilde{K}_2 \cos^2 \varphi + \widetilde{K}_1
 \sin^2 \varphi}{\widetilde{K}_1 \widetilde{K}_2 + \left[
 (\widetilde{K}_1 - \widetilde{K}_2)^2 - (\widetilde{K}_3 +
 \widetilde{K}_4)^2 \right] \sin^2 \varphi \cos^2 \varphi }
\nonumber \\
&=& \int_{0}^{2\pi} \frac{d \varphi}{\pi}\left[ \frac{\widetilde{K}_1 +
\widetilde{K}_2}{\alpha^{+} - \alpha^{-}\cos^2 2\varphi} +
\frac{(\widetilde{K}_2 - \widetilde{K}_1)\cos 2\varphi}
{\alpha^{+} - \alpha^{-}\cos^2 2\varphi} \right]
\nonumber
\end{eqnarray}
where
\begin{equation}
\alpha^{\pm} = (\widetilde{K}_1 \pm \widetilde{K}_2)^2 -
(\widetilde{K}_3+\widetilde{K}_4)^2.
\end{equation}
The second term in the last line of Eq.~(\ref{intL0}) is identically zero and
we are left with
\begin{eqnarray}
\widetilde{L}_0 &=& (\widetilde{K}_1 + \widetilde{K}_2)
\int_{0}^{2\pi} \frac{d \varphi}{\pi} \frac{1}{\alpha^{+} - \alpha^{-}
\cos^2 2\phi} \nonumber \\
&=& 2\frac{\widetilde{K}_1 + \widetilde{K}_2} {\sqrt{\alpha^{+}(\alpha^{+} -
\alpha^{-})}} \nonumber \\
&=& \frac{\widetilde{K}_1+\widetilde{K}_2}{\sqrt{\widetilde{K}_1
\widetilde{K}_2 \left[(\widetilde{K}_1+\widetilde{K}_2)^2
-(\widetilde{K}_3 + \widetilde{K}_4)^2\right]}}, \label{L0dual}
\end{eqnarray}
which can be written in terms of our original coupling constants $K_i$ as
\begin{equation}
\widetilde{L}_0 = \frac{4\pi^2}{p^2T^2}
\frac{K_1(K_2^2-K_3^2)+K_2(K_1^2-K_4^2)}
{\sqrt{K_1K_2\left[(K_1+K_2)^2-(K_3+K_4)^2)\right]}}.
\label{L0dualnt}
\end{equation}
This is not to be confused with $L_0$ from Eq.~(\ref{rgall}) which
is given by
\begin{equation}
L_0 = \frac{K_1+K_2}{\sqrt{K_1K_2\left[(K_1+K_2)^2-(K_3+K_4)^2\right]}}.
\label{L0reg}
\end{equation}

\begin{figure}
\centering
\includegraphics[width=3in]{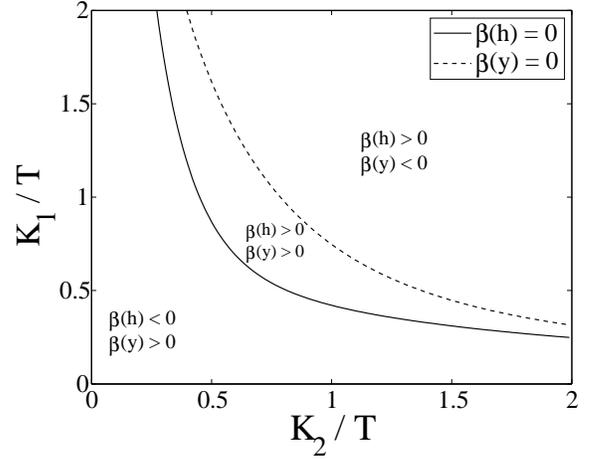}
\caption{The signs of $\beta(h)$ and $\beta(y)$ in the
$K_1-K_2$ plane for $K_3 = -3K_2/5$ and $K_4 = 4K_1/5$. The line
which indicates $\beta(h)=0$ always lies below the line describing $\beta(y)=0$
and consequently $h$ and $y$ are never both irrelevant.}
\label{figbetahy}
\end{figure}

For the direct transition from A to C (Section~\ref{sec:dual}) for $p=4$ we wish
to determine if there is any region close to the manifold of fixed points
where both $h$ and $y$ are irrelevant. We do this by expanding
Eqs.~(\ref{L0dualnt}) and (\ref{L0reg}) for small $K_3$ and $K_4$
parameterized by
\begin{equation}
K_3 = \delta_3 K_2 \quad \mbox{;} \quad K_4 = \delta_4 K_1,
\end{equation}
where $|\delta_3| \mbox{ and } |\delta_4| < 1$ and substituting into
the expressions for $dh/d\ell$ and $dy/d\ell$ in Eq.~(\ref{rgall}).  Define
\begin{equation}
\beta(h) = \frac{1}{h}\frac{d h}{d\ell} \quad \mbox{;} \quad
\beta(y) = \frac{1}{y}\frac{d y}{d\ell},
\end{equation}
it is then straightforward to show that $\beta(h) \gtrless 0$ if
\begin{equation}
K_1 \gtrless K_1^\ast\left[1 + \left(\frac{\pi^2 K_2^2
\delta_3 + 4\delta_4T^2}{\pi^2 K_2^2+4T^2}\right)^2 \right] +
\mathrm{O}\left(\delta_i^3\right)
\label{K1h}
\end{equation}
and $\beta(y) \lessgtr 0$ if
\begin{eqnarray}
K_1 &\gtrless& K_1^\ast \left[1 + \left(\frac{\pi^2 K_2^2
\delta_3 + 4\delta_4T^2}{\pi^2 K_2^2+4T^2}\right)^2 \right. \nonumber \\
&& \left.+ 2\left(\frac{2\pi K_2 T (\delta_4-\delta_3)}
{\pi^2 K_2^2 + 4T^2}\right)^2 \right ] + \mathrm{O}\left(\delta_i^3\right)
\label{K1y}
\end{eqnarray}
where $K_1^\ast = 4T^2/(\pi^2 K_2)$. The relative signs of $\beta(h)$ and
$\beta(y)$ are shown in Fig.~\ref{figbetahy} and from
Eqs.~(\ref{K1h}) and (\ref{K1y}) we have calculated that the two lines
describing the zeros of $\beta(h)$ and $\beta(y)$ are always separated by
\begin{equation}
\frac{32T^4K_2(\delta_4-\delta_3)^2}{(\pi^2K_2^2+4T^2)^2} > 0
\end{equation}
and thus there are no values of $K_1$ and $K_2$ for small $K_3$
and $K_4$ where the $4$-fold anisotropy and vortices are
\emph{both} irrelevant.

\section{Duality mapping}
\label{app:dual}

This appendix will outline the derivation of Eq.~(\ref{dual}).

It is useful to first consider $\mathcal{F}_\theta$ in
Eq.~(\ref{ftheta}) with $h=\eta=0$, but to include the effect of
vortices. In other words, we do wish the impose periodicity under
$\theta_i \rightarrow \theta_i + 2 \pi$.

First, we write the relevant portion of the action in the form
\begin{equation}
\frac{1}{2} \int d^2 r \partial_i \theta_a C_{ia,jb} \partial_j
\theta_b
\end{equation}
where the indices $i$, $j$, $a$, $b$ extend over $x$, $y$, and the
matrix of couplings $C$ can be easily related to the elastic
constants $K_i$ in Eq.~(\ref{ftheta}). Now, we decouple this by a
set of currents $J_{ia}$ and write the action as
\begin{equation}
\int d^2 r \left[ \frac{1}{2} J_{ia} C_{ia,jb}^{-1} J_{jb} + i
J_{ia} \partial_i \theta_a \right]
\end{equation}
Imposing periodicity in $\theta_i \rightarrow \theta_i + 2 \pi$ is
now equivalent to the requirement that the $J_{ia}$ are integers.
We now integrate the $\theta_a$ out, and solve the resulting
constraint equations by writing
\begin{equation}
J_{ia} = \frac{p}{2\pi} \epsilon_{ij} \partial_j
\widetilde{\theta}_a.
\end{equation}
The constraints that the $J_{ia}$ are integers can now be imposed
by demanding that the $\widetilde{\theta}_a$ take values which are
integer multiples of $2\pi/p$. As usual, we can soften this
constraint by introducing a vortex fugacity $y$, and so obtain the
effective action
\begin{equation}
\int d^2 r \left[ \frac{p^2}{8 \pi^2} \epsilon_{ii'} \partial_{i'}
\widetilde{\theta}_a C_{ia,jb}^{-1} \epsilon_{jj'} \partial_{j'}
\widetilde{\theta}_b - y \sum_a \cos (p \widetilde{\theta}_a)
\right]
\end{equation}
Upon explicitly working out the inverse of the matrix $C$, we find
that this action has exactly the same form as the original
$\mathcal{F}_{\theta}$, with the vortex fugacity $y$ playing the
role of the lock-in field $h$, and the couplings $K_i$ replaced by
the couplings $\widetilde{K}_i$ in Eq.~(\ref{kdual}).

\end{document}